\documentclass[aps,pra,twocolumn]{revtex4}
\usepackage{amsmath}
\usepackage{amssymb}

\newcommand\beq{\begin{equation}}
\newcommand\eeq{\end{equation}}
\newcommand\bea{\begin{eqnarray}}
\newcommand\eea{\end{eqnarray}}
\newcommand\nn{\nonumber}

{}

\newcommand\gdet{\sqrt{g}}
\newcommand\Vtrue{{\widetilde V}}
\newcommand\eaA{e^a_A}

\begin{document}
\title{The covariant Langevin equation of diffusion on Riemannian manifolds}    
\author{Lajos Di\'osi}
\email{diosi.lajos@hun-ren.wigner.hu}
\homepage{www.wigner.hu/~diosi} 
\affiliation{Wigner Research Center for Physics, H-1525 Budapest 114 , P.O.Box 49, Hungary}
\affiliation{E\"otv\"os Lor\'and University, H-1117 Budapest, P\'azm\'any P\'eter stny. 1/A}
\date{\today}

\begin{abstract}
The covariant form of the multivariable diffusion-drift process is
described by the covariant Fokker--Planck equation using the standard
toolbox of Riemann geometry. The covariant form of the adapted
Langevin stochastic differential equation is long sought after in both
physics and mathematics.. We  show that the 
simplest covariant Stratonovich stochastic differential equation 
depending on the local orthogonal frame (cf. vielbein)
becomes the desired covariant Langevin equation 
provided  we impose an additional covariant constraint: 
the vectors of the frame must be divergence-free.

\end{abstract}

\maketitle
\section{Introduction}
Differential geometric studies of stochastic differential equations (SDEs) of diffusion
go back to Dynkin \cite{dynkin1968diffusion} and  Ito \cite{ito1975stochastic}. 
Diffusion  on Riemannian manifolds is described by the Fokker--Planck equation (FPE)
and the celebrated Eells-Elworthy-Malliavin construction 
\cite{elworthy1982stochastic,malliavin1976stochastic,hsu2002stochastic}
yields the adapted diffeomorphism invariant stochastic process.
But its  `canonical' SDE ( i.e.: covariant in local coordinates) is notoriously missing.
In physics, Graham studied diffeomorphism invariant diffusion on the thermodynamic state space
in seminal papers
\cite{graham1977path,graham1977covariant,graham1985covariant,
grabert1980fluctuations}, searching in vain for the covariant Langevin SDE.
The need for a solution has recently re-emerged in the theory of hybrid classical-quantum
dynamics \cite{diosi2023hybrid,diosi2023erratum}.
Our work finds a simple construction of the covariant Langevin SDE adapted to the
covariant FPE and we conjecture that it  is the missing `canonical' SDE.

Differently from the typical mathematical literature which defines the Riemannian 
manifold first and introduces the natural diffusion equation on it, here we
define the diffusion equation on a manifold first and impose  the natural Riemannian
structure afterwards. Accordingly, 
Sec. II recapitulates the usual FPEs and Langevin SDEs first, then 
Sec. III adds the Riemannian structure and transforms the FPE and SDE
 into their equivalent covariant forms, utilizing our finding:
 a mandatory covariant constraint on the Langevin SDE's covariant 
parameters. Comparisons with previous mathematical works are postponed to
Sec. IV, also containing our conclusions.      

\section{Fokker--Planck vs Langevin equation}
The most common irreversible phenomena in physics are diffusive ones, modelled
mathematically by the  FPE. If $P(x)$ is the normalized
probability distribution of an abstract particle of coordinates 
$x=(x^1,x^2,\dots,x^n)$ then the FPE reads \cite{gardiner2009stochastic}:
\beq\label{FPE}
\frac{\partial P}{\partial t}=\tfrac12\left(g^{ab}P\right)_{\!,ab}-\bigl(\Vtrue^a P\bigr)_{,a}~.
\eeq
Here $g^{ab}(x)$ is the positive diffusion matrix and $\Vtrue^a(x)$
is the drift, defined respectively by the expectation value of diffusion and velocity
if the particle is at the position $x$:
\bea
\label{diff}
g^{ab}(x)&=&\frac{d}{dt}\langle x^a x^b\rangle_{P(z)=\delta(z-x)}~,\\
\label{Vtrue} 
\Vtrue^a(x)&=&\frac{d}{dt}\langle x^a\rangle_{P(z)=\delta(z-x)}~.
\eea
With one eye on forthcoming considerations of covariance, we  use the
formalism of general relativity: summation of identical labels is understood,
partial derivatives $\partial/\partial x^a$ are denoted by lower label $a$
with the comma.  

The same diffusive phenomena can alternatively be represented by the adapted stochastic
processes $x_t$ satisfying  Langevin SDEs.
The equivalence between the FPE and the SDE means the following relationship:
\beq\label{FPESDE}
P_t(x)=\bigl\langle\delta(x-x_t)\bigr\rangle~,
\eeq
where $\langle\dots\rangle$ stands for averaging over the stochastic $x_t$.
With $n$ independent Wiener processes $W^A$, the Ito form of the 
Langevin SDE  of $x_t$ is the following  \cite{gardiner2009stochastic}:
\beq\label{SDEIto}
dx^a=\eaA dW^A+\Vtrue^a dt~.
\eeq 
Summation from $1$ to $n$ over repeated labels $A$ is understood
and the matrices $\eaA(x)$ satisfy
\beq\label{frame}
\delta^{AB}\eaA e^b_B=g^{ab}~.
\eeq 
This condition allows for a local orthogonal gauge-freedom:
\beq\label{gauge}
\eaA\Rightarrow O_A^B e^a_B
\eeq
with orthogonal matrices  $O_A^B(x)$. The form of the SDE \eqref{SDEIto} is 
gauge-dependent but the stochastic process $x_t$ is unique.

%For completeness, 
%new
Using  the method e.g.  in ref. \cite{diosi2023hybrid},
we verify the relationship \eqref{FPESDE}, 
Suppose it holds at time $t$, then
we have to show that $dP_t(x)=\langle d\delta(x-x_t)\rangle$ is satisfied if
the l.h.s. is given by the FPE \eqref{FPE} and the r.h.s. is
given by the SDE \eqref{SDEIto}. 

Let us workout the r.h.s.:
\bea
&&\frac{1}{dt}\langle d\delta(x-x_t)\rangle=\nn\\
&=&\frac{1}{dt}\langle-\delta_{,a}(x-x_t)dx^a_t+\tfrac12\delta_{,ab}(x-x_t)dx^a_t dx^b_t\rangle=\nn\\
&=&\langle -\delta_{,a}(x-x_t)\Vtrue^a(x_t)+\tfrac12\delta_{,ab}(x-x_t)g^{ab}(x_t)\rangle=\nn\\
&=&-\bigl(\langle\delta(x-x_t)\rangle\Vtrue^a(x)\bigr)_{,a}
      +\tfrac12\bigl(\langle\delta(x-x_t)\rangle g^{ab}(x)\bigr)_{,ab}\nn\\
&=&-\bigl(P(x)\Vtrue^a(x)\bigr)_{,a}
      +\tfrac12\bigl(P(x)g^{ab}(x)\bigr)_{,ab}~.   
\eea
First we calculated $d\delta(x-x_t)$ with the Ito correction, then inserted
$dx^a$ from the SDE \eqref{SDEIto}. Next, we moved derivations in front
of the expressions so that we could replace the argument $x_t$ 
of both $\Vtrue^a$ and of $g^{ab}$ by $x$, thanks to the $\delta$-function. 
Finally we inserted our initial assumption that $\eqref{FPESDE}$ holds at $t$. 
The result coincides with $dP_t(x)/dt$ calculated from the FPE \eqref{FPE}.

\section{Covariance}
Neither the FPE \eqref{FPE} nor the Ito--Langevin SDE \eqref{SDEIto} are
covariant under general transformations of the coordinates $x^a$.
The common reason of their non-covariance is the non-covariance of
the drift vector \eqref{Vtrue}. For example, if the velocity $\Vtrue^a$ 
vanishes in Euclidean  coordinates it becomes non-zero in curvilinear ones.

The desired covariant FPE is easily achieved.  We borrow the toolbox
of Riemann geometry well-known from general relativity \cite{landau2013classical}. 
Accordingly, we impose a Riemann geometry structure on the 
manifold of coordinates $x$ by identifying the diffusion matrix $g^{ab}$
with the contravariant metric tensor and we introduce the scalar 
probability density $\rho=P/\gdet$ of
covariant normalization
\beq
\int\rho(x)\gdet dx = 1~.
\eeq
The  covariant form of the FPE \eqref{FPE} follows:
\beq\label{FPEcov}
\frac{d\rho}{dt}=\tfrac12 g^{ab}\rho_{;ab} -(V^a\rho) _{;a}~,
\eeq
where semicolons denote covariant derivatives and $V^a$ is the co(ntra)variant drift:
\beq\label{Vcov} 
 V^a=\Vtrue^a-\frac{1}{2\gdet}(\gdet g^{ab})_{,b}~.
\eeq
As a price of its covariance, this velocity \emph{parameter} is different from the 
\emph{true} but non-covariant drift velocity $\Vtrue^a$ defined by  \eqref{Vtrue}. 
%The covariant $V^a$ is gauge-dependent. it coincides with
%the true drift $\Vtrue^a$ in the harmonic gauge defined just by 
%$(x^a)_{;bc}g^{bc}=(1/\gdet)(\gdet g^{ab})_{,b}=0$ for each individual $x^a$. 

Now we propose the covariant Langevin equation. The matrix
$\eaA$, introduced for the non-covariant Ito--Langevin SDE \eqref{SDEIto},
is standard in Riemann geometry. It is called \emph{frame} (or vielbein, also tetrad
in the four-dimensional pseudo-Riemann space of general relativity).
The condition \eqref{frame} is called the frame's orthogonality condition.
And now we impose our \emph{new covariant constraint} on the frame. Namely, 
the covariant divergence of the frame's $n$ orthogonal vectors should vanish each:
\beq\label{newconstr}
(\eaA)_{;a}=0~.
\eeq
We mention that the choice of the
frame still has a gauge-freedom which is a restriction of 
\eqref{gauge}, not detailed here.     

The  covariant form of the non-covariant Ito--Langevin SDE \eqref{SDEIto}
is, as we prove below, simple enough:
\beq\label{SDEcov}
dx^a=\eaA\circ dW^A+V^a dt~,
\eeq
where $\circ$ means Stratonovich product instead of Ito's.
The r.h.s.  is explicit covariant. This is compatible with
the  covariance of the l.h.s. since the Stratonovich differentials
satisfy the chain rule exactly like common differentials. 
In our case,  if we change the coordinates for $y^a$ then 
the Stratonovich differentials transform covariantly:
\beq
dy^a=\frac{\partial y^a}{\partial x_b}dx^b~.
\eeq
%Note that the Stratonovich SDE \eqref{SDEcov} does not inherit the gauge-freedom
%\eqref{gauge} from the Ito SDE \eqref{SDEIto}. On the contrary, the local
%orthogonal transformations of the frame may correspond to 
%different stochastic processes, values of $\Vtrue^a$ may be different 
%from the true value \eqref{Vtrue} fixed by both \eqref{FPE} and
%\eqref{FPEcov} in coincidence. The new constraint \eqref{newconstr} makes
%the stochastic process $x_t$ unique, it coincides with the solution of the 
%non-covariant Langevin SDE \eqref{SDEIto} as it should.
%We mention that the choice of the
%frame in \eqref{SDEcov} still has a gauge-freedom which is a subgroup of 
%\eqref{gauge}, not specified here.     

Now we prove  that the covariant Stratonovich--Langevin SDE \eqref{SDEcov} 
is equivalent indeed with the non-covariant SDE  \eqref{SDEIto}.
The Ito form of a Stratonovich SDE, like our \eqref{SDEcov}, reads \cite{gardiner2009stochastic}:
\bea\label{ItofromStr}
dx^a\!\!&=&\!\!\eaA dW^A+\frac12 \delta^{AB}(\eaA)_{,b}e^b_B dt + V^a dt=\\
               &=&\!\!\eaA dW^A+\frac12 \delta^{AB}(\eaA)_{,b}e^b_B dt 
                                          +\!\left(\!\Vtrue^a-\frac{(\gdet g^{ab})_{,b}}{2\gdet}\!\right)\!\!dt\nn.     
\eea
Observe that the new drift term  contains the standard partial
derivatives of the frame, not the covariant ones. We are going to work it out:
\bea
\delta^{AB}(\eaA)_{,b}e^b_B&=&g^{ab}_{,b}-\eaA(e^b_A)_{,b}=\nn\\
                                   &=&g^{ab}_{,b}+\eaA\Gamma^b_{bc}e^c_A=\nn\\
                                   &=&g^{ab}_{,b}+g^{ac}\Gamma^b_{bc}=\nn\\
                                   &=&\frac{1}{\gdet}(\gdet g^{ab})_{,b}~.
\eea
In the four steps we used the orthogonality \eqref{frame} of the frame,
the constraint \eqref{newconstr} on its covariant divergence, then \eqref{frame}
again, and the identity $\Gamma^b_{ab}=(\log\gdet)_{,a}$.
If we insert the result in the SDE \eqref{ItofromStr} we recognize the coincidence
with the non-covariant SDE \eqref{SDEIto}.  %The covariant SDE \eqref{SDEcov}
%yields the true drift \eqref{Vtrue}. It would not do so without the constraint $(\eaA)_{;a}=0$. 

We have not yet asked if divergence-free frames  $\eaA$ exist at all. 
The answer is reassuring \cite{paus1998sum}. They exist --- at least locally --- for $n>2$.
Interestingly enough,  they do not exist for $n=2$ unless the geometry is flat.
Construction of the divergence-free frames is trivial on flat Riemannian manifolds.
Then the coordinates $x^a$ can be functions of Euclidean coordinates $y^A$.
Accordingly,  $x^a=f^a(y)$ and the map from Euclidean to curvilinear 
coordinates satisfies the relationship between the Euclidean $\delta^{AB}$ 
and the curvilinear metric tensors:
\beq
\delta^{AB}f^a_{,A}f^b_{,B}=g^{ab}~.
\eeq
The frame's orthogonality condition \eqref{frame} is then satisfied if 
we choose the frame as 
\beq
\eaA=f^a_{,A}~.
\eeq
This frame is divergence-free.
Indeed, the covariant divergence $(\eaA)_{;a}$ vanishes in any curvilinear
coordinates because it vanishes in the particular Euclidean coordinates
where $\eaA=\delta^a_A$.

\section{Discussion}
The covariant Langevin equation \eqref{SDEcov}, known in itself by both Ito \cite{ito1975stochastic}
and  Graham \cite{graham1977covariant}, is describing a frame dependent stochastic
process generically different from the adapted process to the given covariant FPE \eqref{FPEcov}.
Obviously, the choice of the orthogonal frames must not be left completely free.  
Originally, Ito \cite{ito1975stochastic} proposed that 
the frame follow stochastic parallel transport along the stochastic trajectory $x_t$,
the covariant derivative of the frame along the trajectory
be vanishing. In other words,  
\beq
d\eaA=-\Gamma^a_{bc}e^b_A\circ dx^c, 
\eeq
which is a second SDE coupled to the Langevin equation \eqref{SDEcov} of $dx^a$,
see also eq. (3.3.9) in ref. \cite{hsu2002stochastic}.
With Ito's stochastic parallel transport, the covariant SDE \eqref{SDEcov}
becomes a bit more involved than with our deterministic constraint $e^a_{A;a}=0$
but obtains the same stochastic process $x_t$.
The frame is defined on the stochastic trajectory only,  %Such Langevin SDE
%is useful for Euler fluid equations with fluctuations in Riemann space
%\cite{koide2020variational}. 
differently from our proposal where  the frame $\eaA$ is
a given smooth function on the whole Riemannian  manifold.

%Such Langevin SDE
%is useful for Euler fluid equations with fluctuations in Riemann space
%\cite{koide2020variational}. 
%differently from our proposal where  the frame $\eaA$ is
%a given smooth function on the whole Riemannian  manifold. %Its choice
%has been problematic because the true drift $\Vtrue^a$ depends 
%on the choice. Graham's work \cite{graham1977covariant},
%motivated by transport processes in non-linear and non-equilibrium 
%thermodynamic media, contains the covariant FPE \eqref{FPEcov} and
%the relationship \eqref{Vcov} between the true non-covariant and
%covariant drifts, with the observation that their difference  vanishes in 
%harmonic coordinates.   
%Also 
%the covariant Stratonovich--Langevin SDE \eqref{SDEcov} %is proposed
%but solely with the standard orthogonality constraint \eqref{frame}:
%Ref. \cite{graham1985covariant} already puts the ambiguity \eqref{gauge}
%of the frame under  investigation and observes that it is not gauge
%freedom in the Stratonovich SDE. Abandoning the Stratonovich form, 
%the author proposes covariant Ito differential.    

%In Polettini's work \cite{polettini2013generally}  the proposed covariant Ito SDE
%contains separate Christoffel symbols not those inside covariant derivatives.  
%We mention an other particular from of non-standard covariance
%\cite{graham1977covariant,grabert1980fluctuations,ding2020covariant}.
%If a unique equilibrium state $P^{\mathrm{eq}}(x)$ exists then both
%the both the FPE and the SDE can be parametrized by $P^{\mathrm{eq}}(x)$
%instead of the non-covariant drift ${\widetilde V^a}$.

In the mathematical literature of diffusion on Riemannian manifolds, the lack of
`canonical' SDE is explained by the fact that  the Laplace-Beltrami operator $\Delta_{LB}$
is not a sum of operator squares, 
cf. p 75 in \cite{hsu2002stochastic} or in \cite{inauen2023stochastic} most recently. 
That seems to contradict to the existence
of the covariant Langevin SDE. Fortunately, there is no contradiction.
We can always write $\Delta_{LB}$ as sum of operator squares (at least locally and unless $n=2$):
\beq
\Delta_{LB}\equiv%\frac{1}{\gdet}\partial_a\gdet g^{ab}\partial_b=
                      \nabla_a g^{ab}\nabla_b=
\delta^{AB}%(\eaA\partial_a)(e_B^b\partial_b)~.
                          (\eaA\nabla_a)(e_B^b\nabla_b)~,
\eeq
where $\nabla_a$ means the covariant derivative.
The proof is easy if we substitute $g^{ab}=\delta^{AB}\eaA e_B^b$
and use the constraint  $\nabla_a\eaA=0$. 

In summary, 
we have proved that once the covariant Fokker--Planck equation is given, the long-sought 
stochastic differential equation of the adapted process, i.e.,
the  covariant Langevin equation is the Stratonovich 
stochastic differential equation  \eqref{SDEcov} 
containing covariant objects only, as it should, while
our main result is that the covariant divergence of the orthogonal frame $e^a_A$ must
be set vanishing \eqref{newconstr}. %The Langevin equation in itself, 
%proposed originally by Graham in 1977, would keep yielding incorrect drifts
%otherwise.
Our result can and should be refined by more rigorous methods of probability theory and differential
geometry.

 \begin{acknowledgments}
I thank Laszl\'o B. Szabados for valuable discussions,
T. Koide who draw Ref. \cite{ito1975stochastic}
to my attention, and an anonymous referee of the original manuscript who warned me that the problem 
has been discussed extensively and perhaps  solved  already in the mathematical literature.
This research was funded by the Foundational Questions Institute and Fetzer Franklin
Fund, a donor-advised fund of the Silicon Valley Community Foundation 
(Grant No's. FQXi-RFPCPW-2008,  FQXi-MGA-2103), 
the National Research, Development and Innovation Office (Hungary)
``Frontline'' Research Excellence Program (Grant No. KKP133827),
and the John Templeton Foundation (Grant 62099).
\end{acknowledgments}

\bibliography{diosi2024}{}
\end{document}